\documentclass[seceq]{ptptex}

\usepackage{graphicx}
\usepackage{amssymb,latexsym}

\newcommand{\beq}{\begin{equation}}
\newcommand{\beqa}{\begin{eqnarray}}
\newcommand{\eeq}{\end{equation}}
\newcommand{\eeqa}{\end{eqnarray}}

\newcommand{\siml}{\lesssim}

\def\MNRAS{{\it Mon. Not. R. Astron. Soc. }}

\title{
The Constancy of the Constants of Nature: Updates}

\author{
Takeshi \textsc{Chiba}%
}

\inst{
Department of Physics, College of Humanities and Sciences, Nihon University, 
Tokyo 156-8550, Japan
}

\abst{
The current observational and experimental constraints on  the time 
variation of the constants of nature (the fine structure constant $\alpha$, 
the gravitational constant $G$ and the proton-electron mass ratio $\mu=m_p/m_e$) 
are reviewed. 
}

\begin{document}


\maketitle

\section{Introduction}

The title of this paper is oxymoron. We mean the title as  
the possibility of variation of the quantities which are 
supposed to be constants. 

There are two aspects of the constants of nature: the system of units and 
the laws of nature. 
On the one hand, when we attempt to describe natural phenomena using physics 
laws, physical quantities  are expressed using physical constants: 
the speed of light $c$, Planck (Dirac) constant $h (\hbar)$, 
the gravitational constant $G$, 
Boltzmann constant $k$, electron (proton) mass $m_e (m_p)$, for example. 
When combined these constants with the electric constant $\epsilon_0$, we can 
determine all the units in the SI units.  Thus the physical constants are closely related with  the system of units. 
On the other hand, there are four fundamental forces in nature: 
electromagnetic, weak, strong, gravitational. All the phenomena in nature 
are described by these four forces. The coupling constants which 
describe the strength of these forces are physical constants of fundamental 
importance: the fine structure constant 
$\alpha=e^2/4\pi \epsilon_0 \hbar c\simeq 7.30\times 10^{-3}\simeq 1/137$, the 
Fermi coupling constant $G_F/(\hbar c)^3\simeq 1.17\times 10^{-5}{\rm GeV}^{-2}$, 
the strong coupling constant $\alpha_S\simeq 0.119$, 
the gravitational constant $G\simeq 6.67\times 10^{-11}{\rm m^3kg^{-1}s^{-2}}$  
(a dimensionless gravitational fine structure constant 
$\alpha_G=Gm_p^2/\hbar c\simeq 5.10\times 10^{-39}$ may also be used).  
Therefore, if these constants are not constant, then the correspondence between 
the experimental results and theories would depend on when and where the 
measurements are performed, which would result in the violation of the 
universality of the laws of nature. Testing the constancy of the physical constants 
thus is of fundamental importance. 

\subsection{Large Number Hypothesis}

Dirac appears to have been the first who argued 
for the possibility of  time variation of the constants of nature \cite{dirac37}. 
As is well-known, dimensionless 
numbers involving $G$ are huge (or minuscule). For example, the ratio 
of the electrostatic force to the gravitational force  between an electron 
and a proton is 
\beq
N_1=\frac{e^2}{ Gm_pm_e}\simeq 2\times 10^{39},
\eeq
where $e$ is the electric charge, $m_p$ is the proton mass and $m_e$ is 
the electron mass. 
Similarly, the ratio of the Hubble horizon radius of the Universe, 
$H_0^{-1}$ to the classical radius of an electron is
\beq
N_2=\frac{cH_0^{-1}}{ e^2m_e^{-1}c^{-2}}\simeq 3\times 10^{40}h^{-1},
\eeq
where $h$ is the Hubble parameter in units of $100{\rm km s^{-1}Mpc^{-1}}$.
Curiously, the two nearly coincide, which motivated Dirac to postulate 
the so-called the large number hypothesis \cite{dirac38}. 
In his article entitled ``A new basis for cosmology'', he describes 
\cite{dirac38} 
\begin{quote}
{\it Any two of the very large dimensionless numbers occurring in Nature are
 connected by a simple mathematical relation, in which the coefficients are 
of the order of magnitude unity.}
\end{quote}
Thus if the (almost) equality $N_1={\cal O}(1)\times N_2$ holds 
always, then  $G$ must decrease with time $G\propto t^{-1}$\citen{dirac37}, 
or the fine structure constant, $\alpha$, must increase with time 
$\alpha \propto t^{1/2}$ \citen{gamov67} since $H \propto t^{-1}$.

Nowadays we know that such a huge dimensionless number like $N_1$ is 
related to the gauge hierarchy problem. In fact, the gauge couplings 
 are {\it running} (however, only logarithmically) 
as the energy grows, and all the gauge couplings are believed to unify at 
the fundamental energy scale (probably string scale). 
The fact that $N_1$ nearly coincides with $N_2$ may be just accidental, and 
pursuing the relation between them is numerological speculation (or requires 
anthropic arguments). However, Pandora's box was opened. 
In the following, we mention several motivations for considering the variation 
of the constants of nature. 

\subsection{Newton, Einstein, String}

Space and time in Newtonian mechanics 
are rigid and immutable: the absolute space and time which define the absolute 
inertial frame and exist forever even without matter. 

The concept of space and time in general relativity is different. 
The structure of space and time is affected by the presence of matter and 
thus becomes soft and malleable. However, the laws of physics are kept rigid: 
the equivalence principle fixes locally the laws of physics. 

On the other hand, string theory can be viewed as a framework for 
softening the laws of physics \cite{damour02}.  
In string theory, the coupling constants are determined by the vacuum 
expectation values of some scalar fields and thus they are no longer constant  
at all. The situation is summarized in Table \ref{tab1}. 

\begin{table}
  \begin{center}
  \setlength{\tabcolsep}{3pt}
  \begin{tabular}{c|c c} 
   & spacetime & laws of physics \\ \hline
 Newton & rigid & rigid \\ 
 Einstein & soft & rigid \\ 
 String Theory & soft & soft \\ 
  \end{tabular}
  \end{center}
\caption{Softening of Spacetime and Laws of Physics}
\label{tab1}
\end{table}

String theory is the most promising approach to unify all  fundamental 
forces in nature. It is believed that in string theory all the 
coupling constants and parameters (except the string tension) in nature are 
derived quantities and are determined by the vacuum expectation values of 
the dilaton and moduli. However, no compelling mechanism how and when
to fix the dilaton/moduli is known.

On the other hand, we know that the Universe is expanding. Then it is 
no wonder to imagine the possibility of the time variation of the 
constants of nature during the evolution of the Universe. 

In fact, it is argued that the effective potentials of dilaton or moduli 
induced by nonperturbative effects may exhibit runaway structure; 
they asymptote zero for the weak coupling limit where dilaton becomes minus 
infinity or internal radius becomes infinity and symmetries are restored 
in the limit \cite{dine85,witten00}. Thus it is expected that as these 
fields vary, the natural ``constants'' may change in time 
and moreover the violation of the weak equivalence principle may be induced 
\citen{witten00,damour94} (see also \citen{marciano84,maeda88} 
for earlier discussion). Moreover, the present cosmic acceleration may be induced by a slowly rolling light scalar field (called quintessence). Quintessence can couple to electromagnetic field \cite{carroll98} and/or gravitational field \cite{chiba99} directly unless such couplings are forbidden by some symmetries. The couplings 
could induce the time variation of $\alpha$ and/or $G$. 

Hence, any detection or nondetection of such variations at various cosmological epochs 
could provide useful information about the nature of dilaton/moduli fixing and the coupling of the quintessence field. 

\subsection{Importance as Null Tests}

We should emphasize another important aspect of checking the constancy of the fundamental constants: a null test. It is of fundamental importance to check to what extent the
gravitational force obeys the inverse square law and to what extent the
equivalence principle (the universality of free-fall) holds. Likewise,
it is of fundamental importance to check the constancy of the 
fundamental constants to the ultimate precision. By comparing the 
experimental values at various epochs and positions, we could confirm 
the internal consistency of the foundation of the laws of physics.

\subsection{Use of Cosmology}

Cosmological observations have played important roles in testing the constancy of 
the fundamental constants, which may be evident by writing the time derivative 
in terms of a difference: 
\beqa
\frac{\Delta \alpha}{\alpha\Delta t}.
\eeqa
Therefore, in order to place a strong constraint on the time variability, one needs 
to (1) measure the constant accurately (thereby minimizing $\Delta\alpha/\alpha$) or 
to (2) measure for a long time (larger $\Delta t$). Laboratory precision 
experiments correspond to the former ($\Delta \alpha/\alpha\ll 1$ but 
$\Delta t \sim {\cal O}(1)$ yr), while cosmological observations the latter (
$\Delta t$ as much as 137 Gyr but $\Delta \alpha/\alpha \sim {\cal O}(1)$).

\subsection{Plan of the Paper}

In this article, we review the current experimental (laboratory, 
astrophysical and geophysical) constraints on  
the time variation of the constants of nature. 
In particular, we consider 
$\alpha$ (sec.2), $G$ (sec.3), $m_p/m_e$ (sec.4) and $\Lambda$ (sec.5),  
extending and updating our previous 
review \citen{chiba}. 
More than ten years have passed since our previous review, 
and significant progress has been made in the experimental constraints on 
the variation  (in particular thanks to the release of the WMAP data), 
so it is very timely to update our review. 
See also \citen{uzan03} for recent reviews. 
For earlier expositions, see \citen{dyson72,bt86} for example.
We sometimes use the units of $\hbar =c=1$ 
and assume $H_0=100h {\rm km/s/Mpc}$ with 
$h=0.71$ for the Hubble parameter and $\Omega_M=0.27$ and 
$\Omega_{\Lambda}=0.73$ for the cosmological parameters taken from WMAP results \citen{wmap}.

\section{$\alpha$}

In this section, we review the experimental constraints on the time variation 
of the fine structure constant. The results are summarized in Table \ref{table:alpha}.

\begin{table}
  \begin{center}
  \setlength{\tabcolsep}{3pt}
  \begin{tabular}{|l|c|c|r|} \hline
  &  redshift & $\Delta\alpha/\alpha$ & $\dot\alpha/\alpha({\rm yr}^{-1})$ \\ \hline
  Atomic Clock(${\rm Yb^+}$/${\rm Hg^+}$/H)\cite{peik04} & 0 && $(-0.3\pm 2.0)\times 10^{-15}$\\ \hline
Atomic Clock(${\rm Hg^+}$/${\rm Yb^+}$/H)\cite{fortier07} & 0 && $(-0.55\pm 0.95)\times 10^{-15}$\\ \hline
Atomic Clock(${\rm Sr}$/${\rm Hg^+}$/${\rm Hg^+}$/H)\cite{blatt08} & 0 && $(-3.3\pm 3.0)\times 10^{-16}$\\ \hline
Atomic Clock(${\rm Al^+}/{\rm Hg^+}$)\cite{rosenband08} & 0 && $(-1.6\pm 2.3)\times 10^{-17}$\\ \hline
Atomic Clock($^{162}$Dy/$^{163}$Dy)\cite{cingoz07} & 0 && $(-2.7\pm 2.6)\times 10^{-15}$\\ \hline
   Oklo(Damour-Dyson\cite{damour96}) &  0.16 & $(-0.9\sim 1.2)\times 10^{-7}$ &
 $(-6.7\sim 5.0)\times 10^{-17}$   \\  \hline
   Oklo(Fujii et al.\cite{fujii00})  & 0.16  & $(-0.18\sim 0.11)\times 10^{-7}$ & 
$(0.2\pm 0.8)\times 10^{-17}$\\  \hline
Oklo(Petrov et al.\cite{petrov06}) &0.16  & $(-0.56\sim 0.66)\times 10^{-7}$ & 
$(-3.7\sim 3.1)\times 10^{-17}$\\  \hline
Oklo(Gould et al.\cite{gould06}) &0.16  & $(-0.24\sim 0.11)\times 10^{-7}$ & 
$(-0.61\sim 1.3)\times 10^{-17}$\\  \hline
Re/Os bound\cite{fujii03}& 0.43 &  $(-0.25\pm 1.6)\times 10^{-6}$      &  $(-4.0\sim 2.9)\times 10^{-14}$ \\ \hline
 HI 21 cm\cite{cowie95} & 1.8 & $(3.5\pm 5.5)\times 10^{-6}$ & $(-3.3\pm 5.2)\times 10^{-16}$ \\  \hline
  HI 21 cm\cite{carilli00} & 0.25,0.68 & $<1.7 \times10^{-5}$ &  \\  \hline
  QSO absorption line(SiIV)\cite{cowie95} & $2.67-3.55$ & $<3.5\times 10^{-4}$ &  \\  \hline
QSO absorption line(MM)\cite{webb99} & $0.5-1.6$ & $(-1.09\pm 0.36)\times 10^{-5}$ &  \\  \hline
  QSO absorption line(MM)\cite{webb00} & $0.5-3.5$ & $(-0.72\pm 0.18)\times 10^{-5}$ &  \\  \hline
  QSO absorption line(SiIV)\cite{webb01}& $2.01-3.03$ & $(-0.5\pm 1.3)\times 10^{-5}$ &  \\  \hline
  QSO absorption line(MM)\cite{webb03} & $0.2-3.7$ & $(-0.543\pm 0.116)\times 10^{-5}$ & \\ \hline
 QSO absorption line(MM)\cite{murphy04} & $0.2-4.2$ & $(-0.573\pm 0.113)\times 10^{-5}$ & \\ \hline 
OH\cite{darling04} & $0.247671$ & $(0.51\pm 1.26)\times 10^{-5}$ &  $(-1.7\pm 4.3)\times 10^{-15}$ \\ \hline 
OH \cite{kanekar10} & 0.247 &  $(-3.1\pm 1.2)\times 10^{-6}$ &  
$(1.1\pm 0.4)\times 10^{-15}$\\ \hline
  QSO absorption line(MgII/FeII)\cite{chand} & $0.4-2.3$ &  $(-0.06\pm 0.06)\times 10^{-5}$ & \\ \hline
QSO absorption line(MgII/FeII)\cite{murphy071} & $0.4-2.3$ &  $(-0.44\pm 0.16)\times 10^{-5}$ & \\ \hline
  QSO absorption line(SiIV)\cite{chand2} & $1.59-2.92$ &  $(0.15\pm 0.43)\times 10^{-5}$ & \\ \hline
  QSO absorption line(FeII)\cite{molaro07} & 1.84 & $(5.66\pm 2.67)\times 10^{-6}$ & $(-5.51\pm 2.60)\times 10^{-16}$ \\ \hline
  QSO absorption line(FeII)\cite{molaro07}  & 1.15 & $(-0.12\pm1.79)\times 10^{-6}$ & 
$(0.14\pm 2.11)\times 10^{-16}$\\ \hline
  QSO absorption line(FeII)\cite{chand3} & 1.15 & $(0.5\pm 2.4)\times 10^{-6}$ & 
$(-0.6\pm 2.8)\times 10^{-16}$\\ \hline
 QSO absorption line(FeII)\cite{agafonova11} & 1.58 & $(-1.5\pm 2.6)\times 10^{-6}$ & 
$(1.5\pm 2.7)\times 10^{-16}$\\ \hline
  CMB\cite{martins} & $10^3$  &  $-0.06 \sim 0.01$ & 
$< 5\times 10^{-12}$ \\  \hline
 CMB\cite{menegoni} & $10^3$  &  $-0.013 \sim 0.015$ & 
$< 1\times 10^{-12}$ \\  \hline
  BBN\cite{cyburt05} & $10^{9}$ &  $< 6\times 10^{-2}$ & $< 4.4\times 10^{-12}$ \\  \hline
  \end{tabular}
  \end{center}
\caption{
Summary of the experimental bounds on  the time variation of the fine 
structure constant. $\Delta\alpha/\alpha\equiv 
(\alpha_{\rm then}-\alpha_{\rm now})/\alpha_{\rm now}$.
}
\label{table:alpha}
\end{table}

\subsection{Earth and $\dot\alpha$: Oklo Natural Reactor and Meteorites}

\subsubsection{Oklo Natural Reactor. }

In 1972, the French CEA (Commissariat \`a l'Energie Atomique) discovered 
ancient natural nuclear reactors in the ore body of the Oklo uranium mine 
in Gabon, West Africa. 
It is called the Oklo phenomenon. The reactor operated about 2 billion 
years ago corresponding to the redshift $z\simeq 0.16$ for 
the assumed cosmology ($h=0.71,\Omega_M=0.27,\Omega_{\Lambda}=0.73$).

Shlyakhter noticed the extremely low resonance energy ($E_r=97.3 {\rm meV}$) 
of the reaction
\beq
~^{149}S_m+n\rightarrow ~^{150}S_m+ \gamma, 
\label{reaction}
\eeq
and hence the abundance of $~^{149}S_m$ (one of the nuclear fission 
products of $~^{235}U$) observed at the Oklo can be a good 
probe of the variability of the coupling constants \cite{shlyakhter76}.
The isotope ratio of $~^{149}S_m/~^{147}S_m$ is $0.02$ rather 
than $0.9$ as in natural samarium due to the neutron flux onto $~^{149}S_m$ 
during the uranium fission. 
The neutron-absorption cross section $\sigma(E)$ of 
the reaction Eq. (\ref{reaction}) is well described by the Breit-Wigner formula, 
\beqa
\sigma(E)=\frac{g\pi \hbar^2}{2m_nE}\frac{\Gamma_n\Gamma_{\gamma}}{(E-E_r)^2+\Gamma^2/4},
\eeqa
where $g$ is the statistical factor and $\Gamma=\Gamma_n+\Gamma_{\gamma}$ is the total width in terms of the neutron and the photon widths. 
{}From an analysis of nuclear and geochemical data, the operating conditions 
of the reactor was inferred and the thermally averaged neutron-absorption 
cross section could be estimated. The result was 
$\Delta E_r=E_r^{Oklo}-E_r^0=(-120 \sim 90) {\rm meV}$ \citen{damour96} and 
$\Delta E_r=4\pm 16$ meV \citen{fujii00}. 
On the other hand, from the mass formula of heavy nuclei,  
the change in resonance energy is related to the change in $\alpha$ 
through the Coulomb energy contribution
\beq
\Delta E_r= -1.1\frac{\Delta \alpha}{\alpha} {\rm MeV}.
\label{er-alpha}
\eeq
By estimating the uncertainty in the resonance energy, Shlyakhter
obtained the famous bound $\dot \alpha/\alpha=10^{-17}{\rm yr}^{-1}$. 
Damour and Dyson reanalyzed the data by carefully estimating the uncertainty and 
obtained $\dot \alpha/\alpha=(-6.7\sim 5.0)\times 10^{-17}{\rm yr}^{-1}$ \citen{damour96}. 
Using new samples that were carefully collected to minimize natural 
contamination and also on a careful temperature estimate of the reactors, 
Fujii et al. reached a tighter bound \footnote{They noted that data is 
also consistent with a non-null result: 
$(-4.9\pm 0.4)\times 10^{-17}{\rm yr}^{-1}$, indicating an apparent 
evidence for the time variability. However, from the analysis of the 
isotope compositions of $G_d$, the consistency of the $S_m$ and $G_d$
results  supports the null results.}
$\dot \alpha/\alpha=(0.2\pm 0.8)\times 10^{-17}{\rm yr}^{-1}$ 
\citen{fujii00}.\footnote{Note the plus sign in front of $0.2$ unlike \citen{fujii00} which should be consistent with 
their value of $\Delta E_r$ and the relation Eq.(\ref{er-alpha}). } 

Recently, the use of the Maxwell-Boltzmann distribution for low energy neutron spectrum 
was questioned and it is claimed that the analysis 
of Oklo data by employing a more realistic spectrum including the $1/E$ tail
implies a decrease in $\alpha$, $\Delta\alpha/\alpha\geq 4.5\times 10^{-8}$ 
with the significance being $6\sigma$ \citen{lt04}. 
Full-scale computations of the Oklo reactor using modern method 
of reactor physics, however, show no evidence for such a change: 
$\Delta\alpha/\alpha=(-5.6\sim 6.6)\times 10^{-8}$ \citen{petrov06}; 
$\Delta\alpha/\alpha=(-2.4\sim 1.1)\times 10^{-8}$ \citen{gould06}.\footnote{Note that 
in the authors of \citen{gould06} use the formula Eq.(\ref{er-alpha}) with the opposite sign 
which is corrected here.} 
The discrepancy arises because the reactor model used in \citen{lt04} is an infinite medium 
reactor model and is found to be undercritical if the reactor is made finite.

\subsubsection{Meteorites.}

Another geophysical bound on the variation of $\alpha$ can be obtained from the 
determination of nuclear decay rates using meteoritic data \cite{pd,dyson72}. 
The isotopes which are most sensitive to changes in $\alpha$ are typically those 
with lowest beta-decay $Q$-value, $Q_{\beta}$. The isotope with the smallest 
$Q_{\beta}(=2.66\pm0.02$keV) value is $~^{187}{\rm Re}$ \citen{pd}. 

The present abundances of $~^{187}{\rm Re}$ and $~^{187}{\rm Os}$ are given by 
\beqa
&&  (~^{187}{\rm Re})_0  =  (~^{187}{\rm Re})_i \exp(-\bar{\lambda}(t_0-t_i)),\\
&&  (~^{187}{\rm Os})_0  =  (~^{187}{\rm Os})_i + (~^{187}{\rm Re})_i 
\left(1-\exp(-\bar{\lambda}(t_0-t_i))\right), 
\eeqa
where the subscripts $0$ and $i$ refer to the present and the initial quantities and 
$\bar{\lambda}$ is the time averaged decay constant: 
$\bar{\lambda} =\int^{t_0}_{t_i}\lambda(t)dt/(t_0-t_i)$. $(~^{187}{\rm Re})_i$ 
can be eliminated to give
\beq
(~^{187}{\rm Os})_0  =  (~^{187}{\rm Os})_i + (~^{187}{\rm Re})_0
\left(\exp(\bar{\lambda}(t_0-t_i))-1\right), 
\eeq
which provides a linear relation (an isochron) between the present abundances 
(relative to $~^{188}{\rm Os}$) of 
$~^{187}{\rm Re}$  and $~^{187}{\rm Os}$. The slope of the linear curve determines 
$\bar{\lambda}$ once the age $t_0-t_i$ is independently determined.  

The $~^{187}{\rm Re}$ decay constant has been determined through the generation 
of high precision isochrons from material of known ages, particularly iron meteorites. 
Using the Re-Os ratios of IIIAB iron meteorites that are thought to have been formed in the
early crystallization of asteroidal cores, Smoliar et al. found a $~^{187}{\rm Re}$ 
decay constant of $\bar{\lambda}=(1.6666\pm 0.009)\times 10^{-11}{\rm yr}^{-1}$ 
assuming that the age of the IIIA iron meteorites is 
$4.5578{\rm Gyr}\pm 0.4{\rm Myr}$ which is identical to the Pb-Pb 
age of angrite meteorites \cite{smoliar} \footnote{Note that the authors of \citen{smoliar} 
recommend the value of $\lambda=(1.6666\pm 0.017)\times 10^{-11}{\rm yr}^{-1}$ 
considering the systematic error associated with the spike calibration 
(see Ref.16 in \citen{smoliar}).}.

The beta-decay constant depends on $Q_{\beta}$ as $\lambda \propto Q_{\beta}^{2.835}$ 
\citen{dyson67}, and if we assume that the variation of $Q_{\beta}$ comes entirely from the 
Coulomb term, we find using the nuclear mass formula \cite{mr} that 
$\Delta Q_{\beta}=-19{\rm MeV}{\Delta\alpha/\alpha}$. Hence, 
\beqa
\frac{\Delta\lambda}{\lambda}=-2.0\times 10^{4}\frac{\Delta\alpha}{\alpha}.
\eeqa
The bound on $\Delta\alpha/\alpha$ 
over the age of the solar system $\simeq 4.6$Gyr ($z\simeq 0.43$) is thus obtained from 
the comparison of the $^{187}{\rm Re}$ meteoritic measurements of the time averaged 
$^{187}{\rm Re}$ decay rate with the recent laboratory measurements 
\cite{decayrate} : 
$\Delta\alpha/\alpha =(2.5\pm 16)\times 10^{-7}$ at $z\simeq 0.43$ \citen{fujii03}.\footnote{
We take the recommended value of $\lambda$ for the meteorites. 
Even if we assume 0.5\% error for the meteoritic measurements, the bound only becomes 
$\Delta\alpha/\alpha=(-2.5\pm 15)\times 10^{-7}$.} 
The difference between \citen{fujii03} and
\citen{olive} comes from the difference in the adopted laboratory
measurements of the decay rate.  We use a more recent measurement \citen{decayrate}. 
It is to be noted, however,  that the bound
depends on the way of time dependence of the $^{187}{\rm Re}$ decay constant
\cite{fujii03}.

\subsection{(Hyper)Fine splitting and $\dot\alpha$}

According to the Dirac equation, the energy levels of a hydrogen-like 
atom with atomic number $Z$ are given by
\beqa
E_{nj}&=&\frac{m_ec^2}{\sqrt{1+Z^2\alpha^2/(n-\delta_j)^2}}\nonumber\\
&=&m_ec^2-\frac{m_ec^2Z^2\alpha^2}{2n^2}-\frac{m_ec^2Z^4\alpha^4}{n^3(2j+1)}+
\frac{3m_ec^2Z^4\alpha^4}{8n^4}+\dots,
\eeqa
where $\delta_j=j+\frac12-\sqrt{(j+\frac12)^2-Z^2\alpha^2}$, $n$ is the principal quantum number and $j$ is the quantum number associated with the total electron angular momentum. The fine structure, which arises due to the spin-orbit coupling, is the difference in energy between levels of different $j$ for the same $n$. 
For example, for the hydrogen atom, $E(2P_{3/2})-E(2P_{1/2})
\simeq m_ec^2\alpha^4/32\simeq 4.53\times 10^{-5}{\rm eV}\simeq hc/(2.75{\rm cm})$. 
The energy levels are further split into doublets (hyperfine structure) 
by the coupling of the proton spin with 
the total electron angular momentum. For example, for the hydrogen atom, the hyperfine splitting for 
$s$ states ($n=1,j=1/2$) is given by \cite{iz}
$\Delta E_{hf}=\frac43m_ec^2\alpha^4\frac{m_e}{m_p}g_p\simeq 5.89\times 10^{-6}{\rm eV}\simeq hc/(21.1 {\rm cm})$, 
where $g_p$ is the proton gyromagnetic ratio. 

Since the fine structure levels depend on $\alpha$, 
the wavelength spectra of cosmologically distant quasars provide a natural 
laboratory  for investigating the time variability of $\alpha$. 
Narrow lines in quasar spectra are produced by absorption of radiation in 
intervening clouds of gas, many of which are enriched with heavy elements. 
Because quasar spectra contain doublet 
absorption lines at a number of redshifts, it is possible to check for 
the time variation of $\alpha$ simply by looking for changes in the doublet 
separation of alkaline-type ions with one outer electron as a function of 
redshift \cite{bahcall65,wolfe76}. 
By looking at Si IV doublet, Cowie and Songaila obtained the constraint up to 
$z\simeq 3$: $|\Delta \alpha/\alpha| < 3.5\times 10^{-4}$ \citen{cowie95}.
Also by comparing the hyperfine 21 cm HI transition with optical atomic 
transitions in the same cloud at $z\simeq 1.8$, 
they obtained a bound on the fractional change in $\alpha$ up to 
redshift $z\simeq 1.8$: 
$\Delta \alpha/\alpha =(\alpha_{z=1.8}-\alpha_0)/\alpha= (3.5\pm 5.5)\times 10^{-6}$, 
corresponding to 
$\dot\alpha/\alpha   =(-3.3\pm 5.2)\times 10^{-16}{\rm yr}^{-1}$ 
\citen{cowie95}. Recently, by comparing the absorption by the HI 21 cm 
hyperfine transition  (at $z=0.25,0.68$) with the absorption by 
molecular rotational transitions,  Carilli et al. obtained a bound: 
$|\Delta \alpha/\alpha| < 1.7\times 10^{-5}$ \citen{carilli00}. 

Webb et al. \cite{webb99} introduced a new technique (called many-multiplet 
method) that compares the absorption wavelengths of magnesium 
and iron atoms in the same absorbing cloud, which is far more 
sensitive to a change in $\alpha$ than the alkaline-doublet method. 
They observed a number of intergalactic clouds at redshifts from 0.5 to 1.6. 
For the entire sample (30 absorption systems of FeII,MgI and MgII) they find 
$\Delta \alpha/\alpha = (-1.09\pm 0.36)\times 10^{-5}$, deviating from zero 
at the 3 $\sigma$. They noted that the deviation is dominated 
by measurements at $z>1$, where 
$\Delta \alpha/\alpha = (-1.9\pm 0.5)\times 10^{-5}$. 

Moreover, Webb et al. \citen{webb00} presented further evidence for the time 
variation of $\alpha$ by reanalyzing the previous data and including new 
sample of Keck/HIRES (High Resolution Echelle Spectrometer) absorption systems. The results indicate a smaller 
$\alpha$ in the past and the optical sample (72 systems 
of MgI,MgII,AlII,AlIII,SiII,CrII,FeII,NiII and ZnII) shows a 
4 $\sigma$ deviation for $0.5 < z < 3.5$: 
$\Delta \alpha/\alpha = (-0.72\pm 0.18)\times 10^{-5}$. 
They noted that the potentially 
significant systematic effects only make the deviation significant. 

The latest analysis of the third sample including 128 absorption systems 
for $0.2<z<3.7$ gives 
$\Delta \alpha/\alpha = (-0.543\pm 0.116)\times 10^{-5}$ 
\citen{webb03}, and the sample is slightly updated to 143 absorption systems 
in \citen{murphy04} to yield $\Delta \alpha/\alpha = (-0.573\pm 0.113)\times 10^{-5}$ for $0.2<z<4.2$. 
Again it is consistent with a smaller 
$\alpha$ in the past. The significance is now 4.7 $\sigma$.

\begin{figure}
\includegraphics[width=13cm]{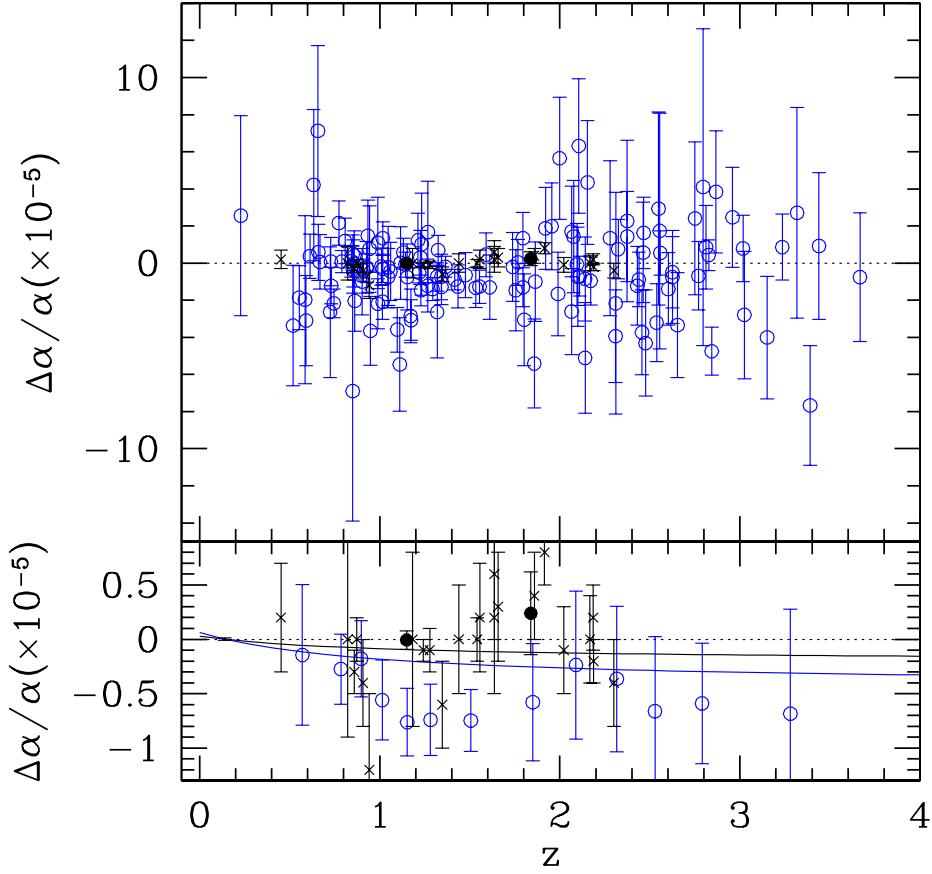}
\caption{The fine structure constant determined by quasar absorption lines 
at several redshifts. Open (blue) circles are the data from \citen{webb99,webb00,webb03}(Keck)
and crosses are from \citen{chand} (VLT)
filled circles are from \citen{levshakov05,levshakov06} (VLT). The lower panel is 
the binned data of \citen{webb99,webb00,webb03} and the value of $\Delta\alpha/\alpha$ 
in each bin is the weighted mean 
with its associated 1$\sigma$ error bar. 
The datum at $z\simeq 0.14$ is the Oklo bound \citen{damour96}. 
A curve is a linear (in scale factor) fit to the data: 
$\Delta\alpha/\alpha=2.93\times10^{-7}-2.31\times10^{-6}(1-a)$ . 
A blue curve is a fit using \citen{murphy071} instead of \citen{chand} :  
 $\Delta\alpha/\alpha=6.08(\pm 2.22)\times10^{-7}-4.85(\pm 1.46)\times10^{-6}(1-a)$.}
 \label{fig:qso}
\end{figure}

More recently, Webb et al. analyzed the dataset from the ESO Very
Large Telescope (VLT) and found the opposite trend: $\alpha$ was {\it
larger} in the past \cite{webb10}.  Combined with the Keck samples, they
claimed the {\it spatial} variation of $\alpha$ \citen{webb10} :
\beqa
\frac{\Delta\alpha}{\alpha}=(1.10\pm0.25)\times 10^{-6} (r/{\rm Glyr})\cos\theta,
\label{alpha:space}
\eeqa
where $r$ is the look-back time $r=ct(z)$ and $\theta$ is the angle
between the direction of the measurement and the axis of best-fit
dipole.

If these observational results are correct, it would have profound 
implications for our understanding of fundamental physics.
So the claim needs to be verified independently by other observations. 
However, recent observations from VLT/UVES (Ultraviolet and Visual Echelle Spectrograph) 
using the same MM method (but only single species) have 
not been able to duplicate these results: for one group, 
$\Delta \alpha/\alpha = (-0.06\pm 0.06)\times 10^{-5}$ 
for MgII/FeII systems at $0.4<z<2.3$ \citen{chand} and 
$\Delta \alpha/\alpha = (0.15\pm 0.43)\times 10^{-5}$ 
for SiIV systems at $1.59\leq z\leq 2.92$ \citen{chand2}, 
while for another group, 
$\Delta \alpha/\alpha = (-0.04\pm 0.46)\times 10^{-5}$ for FeII systems at 
$z=1.15$ \citen{levshakov04}. Recently, in order to avoid the influence of 
spectral shifts due to ionization inhomogeneities in the absorbers and non-zero 
offsets between different exposures, a new method of probing variability of 
$\alpha$ using pairs of FeII lines 
observed in individual exposures (called SIDAM, for single ion differential $\alpha$ 
measurement procedure) is proposed \cite{levshakov05,levshakov06}. 
Using this method, a tighter bound is obtained: for FeII systems at $z=1.839$,  
$\Delta \alpha/\alpha = (5.66\pm 2.67)\times 10^{-6}$ \citen{molaro07} 
and $\Delta \alpha/\alpha = (-0.12\pm 1.79)\times 10^{-6}$ for FeII at 
$z=1.15$ \citen{molaro07}. The data taken by another spectrograph HARPS mounted on 
VLT yield a similar bound: 
$\Delta\alpha/\alpha=(0.05\pm 0.24)\times 10^{-5}$ for the same 
FeII systems at $z=1.1508$ \citen{chand3}. The analysis of 
FeII lines at $z=1.58$ yields  
$\Delta \alpha/\alpha = (-1.5\pm 2.6)\times 10^{-6}$ \citen{agafonova11}. 


It is to be noted, however, that the analysis by Srianand et 
al. \citen{chand} seems suffer from several flaws \cite{fujii} : only about half of 
the observations analyzed in Chand et al. \cite{chand} have
calibration spectra taken before and after the object exposure, 
although in their paper they mentioned that procedure has been 
followed for all the spectra. 
Moreover, the uncertainty in wavelength calibration in \citen{chand} may not be  
consistent with the error in $\Delta\alpha/\alpha$ \citen{levshakov05}. 
According to the analysis of the fundamental noise limitation 
\cite{levshakov05}, the systematic errors in \citen{chand} may be several times 
underestimated. Recent detailed re-analysis of Srianand et al. and Chand et al. 
confirms these concerns: flawed parameter estimation methods in a $\chi^2$ 
minimization analysis \cite{murphy071} (see however, \citen{srianand07})  
and systematic errors in the UVES wavelength calibration \cite{murphy07} \footnote{
Recently, systematic errors in the absolute wavelength calibration of the optical spectrum of HIRES are identified  in \citen{griest10} with typical amplitudes being $\pm 250{\rm m/s}$. 
However, their effects on $\Delta\alpha/\alpha$ are found to be relatively small \cite{murphy09}: 
from $(-0.57\pm 0.11)\times 10^{-5}$ to $(-0.61\pm 0.11)\times 10^{-5}$. }. 
A revised value correcting for respective error is 
$\Delta \alpha/\alpha = (-0.44\pm 0.16)\times 10^{-5}$ ($\chi^2$ minimization) \cite{murphy071} and 
$\Delta \alpha/\alpha = (-0.17\pm 0.06)\times 10^{-5}$  (wavelength calibration) \cite{murphy07}.

The data taken from Keck and VLT are shown in Fig. \ref{fig:qso}. 
A curve is a linear (in scale factor) fit to the data: 
$\Delta\alpha/\alpha=2.93(\pm 1.80)\times10^{-7}-2.31(\pm 1.01)\times10^{-6}(1-a)$ 
and the reduced $\chi^2$ is $\chi^2/{\rm d.o.f}=1.15$  
\footnote{If we use \citen{murphy071} instead of \citen{chand}, the corresponding fitting 
curve is $\Delta\alpha/\alpha=6.08(\pm 2.22)\times10^{-7}-4.85(\pm 1.46)\times10^{-6}(1-a)$ 
(blue curve in lower panel of Fig. 1.)
and the reduced $\chi^2$ is $\chi^2/{\rm d.o.f}=1.53$.}.

We have observed the same objects as Webb et al.'s 
group by Subaru telescope in August 2004 \citen{naoto}. The analysis 
of our observations would provide 
another independent useful information and help to clarify the situation.

\subsection{Laboratory Tests: Clock Comparison}

Laboratory experiments place constraints on the 
present day variation of $\alpha$ and are repeatable and systematic uncertainties 
can be studied by changing experimental conditions, and hence 
such laboratory experiments are 
complementary to the geophysical or cosmological measurements. 
The laboratory constraints so far are based on comparisons of atomic clocks with ultrastable 
oscillators of different physical makeup such as the superconducting 
cavity oscillator vs. the cesium hyperfine clock transition \cite{turneaure76} 
or the Mg fine structure transition vs. the cesium hyperfine clock transition 
\cite{godone93}. In SI units, the second is defined as 
"the duration of 9192631770 periods of the radiation corresponding to the 
transition between the two hyperfine levels 
of the ground state of the $~^{133}{\rm Cs}$  atom" \cite{si98}. 
A cesium atomic clock is the apparatus which tunes the microwave oscillator 
to the same frequency as the resonant absorption frequency of cesium (9192631770Hz). 
Such a clock comparison can be a probe of the time variation of $\alpha$ since 
a hyperfine splitting is a function of $Z\alpha$ ($Z$ is an atomic number) and is 
proportional to $Z\alpha^2(\mu_N/\mu_B)(m_e/m_p)R_{\infty}F_{rel}(Z\alpha)$ 
(where $F_{rel}(Z\alpha)$ is the relativistic correction factor, 
$\mu_N$ is the nuclear magnetic moment, 
$\mu_B=e\hbar/2m_pc$ is the nuclear magneton, and $R_{\infty}=\alpha^2m_ec/4\pi \hbar$ is 
the Rydberg constant). 
More than ten years ago, comparisons of rates between clocks 
based on hyperfine transitions in alkali atoms with 
different atomic number $Z$ (H-maser and 
${\rm Hg^{+}}$ clocks) over 140 days yielded a bound on $\dot\alpha$:  
$|\dot\alpha/\alpha| \leq 3.7\times 10^{-14}{\rm yr}^{-1}$ 
\citen{prestage95}. 

Recently, by comparing a $~^{199}{\rm Hg^{+}}$ optical 
clock ((${\rm ^2S_{1/2} F= 0}$) - (${\rm ^2D_{5/2} F = 2, m_F = 0}$) electric-quadrupole 
transition at 282 nm)  with  
a $~^{133}{\rm Cs}$ clock over 2 years, a much severer upper 
bound has been obtained: $|\dot\alpha/\alpha| \leq 1.2\times 
10^{-15}{\rm yr}^{-1}$ \citen{bize03}. The electric-quadrupole  transition of 
$~^{199}{\rm Hg^{+}}$  is expressed as  $R_{\infty}F_{\rm Hg}(\alpha)$, where 
$F_{\rm Hg}(\alpha)$ is a dimensionless function of $\alpha$. 
Most recent measurement over 6 years gives 
$|\dot\alpha/\alpha| \leq 1.3\times 
10^{-16}{\rm yr}^{-1}$ \citen{fortier07} for $~^{199}{\rm Hg^{+}}$. 
Moreover, the comparison of 
the hyperfine frequencies of $~^{133}{\rm Cs}$ and $~^{87}{\rm Rb}$ atoms 
over nearly 5 years yields $\dot\alpha/\alpha = (0.045\pm 1.6)\times 
10^{-15}{\rm yr}^{-1}$ \citen{marion03}. 
The comparison of the absolute 1S-2S transition in atomic hydrogen to the 
ground state of cesium combined with the results of \citen{bize03,marion03} 
yields a constraint on $\dot\alpha$: $\dot\alpha/\alpha = (-0.9\pm 2.9)\times 
10^{-15}{\rm yr}^{-1}$ \citen{fischer04}. 
However, hyperfine frequencies are sensitive not only to $\alpha$ but also to 
a variation of the nuclear magnetic moment. Moreover, since a microwave distorts 
atoms, the improvements in microwave cesium clocks beyond $10^{-16}$ are unlikely.  
With these motivations, recently, an optical electric quadrupole 
transition frequency at 436 nm in $~^{171}{\rm Yb}^+$ has been measured with a cesium atomic clock 
at two times separated by 2.8 years \cite{peik04}. 
Combined with the data with those for optical transition frequencies in 
$~^{199}{\rm Hg^{+}}$ from \citen{bize03} and in hydrogen from \citen{fischer04} gives 
$\dot\alpha/\alpha = (-0.3\pm 2.0)\times 10^{-15}{\rm yr}^{-1}$ \citen{peik04}. 
Comparisons of ${\rm Hg^+}$ clocks \cite{fortier07} with ${\rm Yb}^+$ and H yield 
$\dot\alpha/\alpha = (-0.55\pm 0.95)\times 10^{-15}{\rm yr}^{-1}$ \citen{fortier07}. 
Also, comparisons of optical Sr clocks \cite{blatt08} with ${\rm Hg^+}$ \cite{fortier07}, 
${\rm Yb^+}$ \cite{peik06}, and H \cite{fischer04} give 
$\dot\alpha/\alpha = (-3.3\pm 3.0)\times 10^{-16}{\rm yr}^{-1}$ \citen{blatt08}. 
Recently, two optical clocks using ${\rm Al^+}$ ((${\rm ^1S_0}$) - (${\rm ^3P_0}$) clock 
transition at 267 nm) \cite{rosenband07} and ${\rm Hg^+}$ are compared directly 
without a cesium atomic clock \cite{rosenband08}. The transition frequency depends  both on 
the Rydberg constant and on $\alpha$ and can be expressed as $R_{\infty}F_{\rm Al}(\alpha)$. 
{}From the ratio of the two transition frequencies, 
 $\dot\alpha/\alpha=(-1.6\pm 2.3)\times 10^{-17}{\rm yr}^{-1}$ is 
obtained \cite{rosenband08}, being independent of the assumptions on the constancy of 
other constants. The frequency uncertainties of the optical clocks are currently  
less than $2.3\times 10^{-17}$.  The accuracy could soon compete  with the gravitational redshifts 
due to the difference in the heights of the clocks ($g\Delta h/c^2 \simeq 10^{-18}(\Delta h/{\rm 1cm})$) so that the optical clocks could be used to map 
the gravitational potential of the earth and to test gravitational physics \cite{rosenband08,chou10}.

More recently, following the proposal of \citen{dzuba03}, 
it is demonstrated that, instead of comparing atomic-clock of different atomic number,   
the difference of the electronic energies of 
the opposite-parity levels in two isotopes of the same atomic dysprosium (Dy) can be 
monitored directly using a radio-frequency electric-dipole transition 
between them \cite{cingoz07}. 
Eight months measurements of the 3.1-MHz transition in $~^{163}{\rm Dy}$ 
and the 235-MHz transition in $~^{162}{\rm Dy}$ show that the 
frequency variation is $9.0\pm 6.7$Hz/yr, $-0.6\pm 6.5$ Hz/yr, respectively, which 
corresponds to 
$\dot\alpha/\alpha = (-5.0\pm 3.7)\times 10^{-15}{\rm yr}^{-1}$ 
for the 3.1-MHz transition and 
$\dot\alpha/\alpha = (-0.3\pm 3.6)\times 10^{-15}{\rm yr}^{-1}$ 
for the 235-MHz transition. The difference frequency gives finally
$\dot\alpha/\alpha = (-2.7\pm 2.6)\times 10^{-15}{\rm yr}^{-1}$ \citen{cingoz07}. 
A unique aspect of this measurement is that the interpretation does 
not require comparison with different measurements to eliminate 
dependence on other constants. Current systematic uncertainties are 
at 1 Hz-level, but mHz-level sensitivity ($|\dot\alpha/\alpha| \sim10^{-18}{\rm yr}^{-1}$) 
may be feasible with this method. 

\subsection{Cosmology and $\dot\alpha$: Big Bang Nucleosynthesis and 
Cosmic Microwave Background}

\subsubsection{Big Bang Nucleosynthesis.}

The process of the Big Bang nucleosynthesis proceeds as follows. 
When the temperature of the universe is greater than 1MeV,  protons and 
neutrons are interchanged by the weak interaction. 
The  neutron-to-proton number ratio $(n/p)$ is given by the equilibrium condition:
\beq
(n/p)= \exp(-Q/T),
\label{yp}
\eeq  
where $Q=1.29$MeV is the mass difference between neutron and proton. 
The equilibrium is violated when the expansion rate of the universe 
$H\simeq \sqrt{G}T^2$ becomes faster than the reaction rate of the weak interaction 
$n\sigma v\simeq G_F^2T^5$, where $G_F$ is the Fermi constant. 
The balance of these two rates determines the freeze-out temperature, 
\beq
T_f \simeq G_F^{-2/3}G^{1/6}\simeq 1{\rm MeV}.
\label{freeze}
\eeq
For the temperature below 1MeV, the number of neutrons decreases only due to the 
natural decay with the life time being 15 minutes. 
Deuterons, ${\rm ^3H_e}$s and finally ${\rm ^4 H_e}$s are produced 
by nuclear interactions of protons and neutrons at $T \simeq 0.1{\rm MeV}$ 
(the first three minutes). 
All the neutrons are incorporated 
into  ${\rm ^4H_e}$ and  the abundance of ${\rm ^4H_e}$, $Y_p$, is given by 
$Y_p=2(n/p)/[1+(n/p)]$.  
Changes in $Y_p$ are induced by changes in $T_f$ and $Q$. However, it 
is found that $Y_p$ is  most sensitive to changes in $Q$ \cite{kolb86}.
The $\alpha$ dependence of $Q$ can be written as 
\cite{gasser,dixit88,campbell95}
\beq
Q \simeq 1.29 -0.76\times \Delta\alpha/\alpha ~~{\rm MeV},
\eeq
and a change in $Y_p$ is related to a change in $\alpha$ as
\beq
\frac{\Delta Y}{Y}\simeq -\frac{\Delta Q}{Q}
\simeq 0.6\frac{\Delta\alpha}{\alpha}.
\label{ypqalpha}
\eeq
Comparing with the observed $Y_p$($Y_p=0.249\pm0.009$)\cite{olive04}, 
a constraint on $\Delta\alpha/\alpha$ is 
obtained: $|\Delta\alpha/\alpha|\leq 6\times 10^{-2}$ \citen{cyburt05}. 
A similar analysis yields a bound on $\Delta Q$: $-4\times 10^{-2}\leq \Delta Q/Q \leq 2.7\times 10^{-2}$ 
\citen{coc07}, which can be translated into a bound on $\Delta\alpha$ via Eq. (\ref{ypqalpha}) as, 
$-4.5\times 10^{-2}\leq \Delta\alpha/\alpha \leq 6.7\times 10^{-2}$.

\subsubsection{Cosmic Microwave Background.}

Changing $\alpha$ changes 
the Thomson scattering cross section, $\sigma_T=8\pi \alpha^2/3m_e^2$, and 
also changes the differential optical depth $\dot\tau$ of photons due to Thomson 
scattering through $\dot\tau =x_en_p\sigma_T$, where $x_e$ is the ionization 
fraction and $n_p$ is the number density of electrons. From the Saha equation, 
 the equilibrium ionization fraction $x^{EQ}_e$ is proportional 
to $(m_e/T)^{3/2}\exp(-\alpha^2m_e/2T)$. Therefore, 
changing $\alpha$ alters the ionization history of the universe and hence 
affects the spectrum of cosmic microwave background fluctuations. 

The last scattering surface is defined by the peak of  
the visibility function, $g(z)=e^{-\tau(z)}d\tau/dz$, which measures 
the differential probability 
that a photon last scattered at redshift $z$. 
As explained in \citen{hannestad99}, increasing $\alpha$ affects the 
visibility function $g(z)$: it increases the redshift of the last scattering 
surface and decreases the thickness of the last scattering surface. 
This is because the equilibrium ionization fraction $x_e^{EQ}$, which is exponentially 
sensitive to $\alpha$, is shifted to 
higher redshift (the effect of increase of $\dot\tau$ due to the increase of $\sigma_T$ 
is minor) and because $x_e$ more closely tracks $x_e^{EQ}$ for larger 
$\alpha$.

An increase in $\alpha$ changes the spectrum of CMB fluctuations: the peak 
positions in the spectrum shift to higher values of $\ell$ (that is, 
a smaller angle) and the values of $C_{\ell}$ (the angular power spectra 
of temperature anisotropies) increase \cite{hannestad99}. 
The former effect is due to  the increase of the redshift of the last 
scattering surface, while the latter is due to a larger early integrated 
Sachs-Wolfe effect because of an earlier recombination. 
Beyond the first peak, the diffusion damping of CMB fluctuation due to the thickness of 
the last scattering surface becomes important \cite{hu96,hannestad99}. 
The diffusion damping is caused by the random walk of 
CMB photons and hence the diffusion length $\lambda_D$ is given by $\lambda_D\simeq 1/\sqrt{H\dot\tau}$. The damping factor of CMB fluctuations is estimated 
as $\sim \exp(-\lambda_D^2/\lambda^2)$ for a given wavelength $\lambda$ of the fluctuations. 
A large $\alpha$ shortens the diffusion length $\lambda_D$ and hence weakens 
the effect of diffusion damping and makes the values of $C_{\ell}$ increase.

The analysis of the first-year observations of CMB fluctuations from  the 
WMAP(Wilkinson Microwave Anisotropy Probe) satellite 
\cite{wmap} gives $-0.06<\Delta\alpha/\alpha<0.01$  at 2 $\sigma$ \citen{martins}. 
The five-year WMAP data combined with the CMB data sets by ACBAR, QUAD, BICEP, 
BOOMERanG and CBI and with the recent measurement of the Hubble constant by HST \cite{hst} 
improves the accuracy:  $-0.011<\Delta\alpha/\alpha<0.015$ at 2 $\sigma$ \citen{menegoni}.
A recent analysis of the seven-year WMAP data combined with the matter power spectrum of 
Sloan Digital Sky Survey LRG yields  $\Delta\alpha/\alpha=-0.014\pm 0.014$ at 2 $\sigma$ \citen{landau10}.

\subsubsection{Constraints at $30<z<1000$ and $z<1$.} 
Several probes of variations in $\alpha$ after the epoch of last scatter have been 
proposed: 21 cm absorption of CMB \cite{wandelt07} and peak luminosity of type Ia supernovae 
(SNIa) \cite{chiba03}. The former probes the variation for redshifts in the range $30<z<1000$, 
while the latter at $z<1$. 

After recombination ($z\sim 1000$) and before reionization ($z\sim 30$),  hydrogen atoms are 
in ground state which is split into a singlet and a triplet state due to hyperfine splitting. 
The absorption of CMB at 21 cm hyperfine transition of the neutral atomic hydrogen is very 
sensitive to the variations in $\alpha$. The Einstein $A$ coefficient of the spontaneous emission 
of the 21 cm transition is proportional to $\alpha^{13}$ and 
the brightness temperature signal of CMB at 21 cm $T_b$ is proportional to $\alpha^5$ 
\citen{wandelt07}. Future radio telescopes may give a constraint on $\Delta \alpha$ of $1\%$ 
\citen{wandelt07}.

A Type Ia supernova is considered to be a good standard candle, because 
its peak luminosity correlates with the rate of decline of the magnitude. 
Observations of Type Ia supernovae have been used to constrain  
cosmological parameters. The homogeneity of the peak 
luminosity is essentially due to the homogeneity of the progenitor mass, and 
this is primarily determined by the Chandrasekhar mass, which is 
proportional to $ G^{-3/2}$. 
The peak luminosity also depends on the diffusion time of photons,  
which depends on $\alpha$ through the opacity. A decrease in opacity reduces 
the diffusion time, allowing trapped radiation to escape more rapidly,
leading in turn to an increase in the luminosity. 
Decreasing $\alpha$
causes the opacity to decrease, which allows photons to escape more 
rapidly, thereby leading to an increase in the luminosity. 
Thus a smaller (larger) value of $\alpha$ would make supernovae 
brighter (fainter). 
The change in the absolute magnitude $\Delta
{\cal M}$ at the peak luminosity is related to the variation in $\alpha$ as
$\Delta {\cal M}\simeq (\Delta\alpha/\alpha)^{-1}$ \citen{chiba03}. 
Future experiments to observe distant SNIa like SNAP would reduce systematic errors 
to a magnitude of 0.02 mag, which corresponds to $\Delta\alpha/\alpha < 2\times 10^{-2}$. 
This bound is significantly larger than the current bound by QSO for redshifts 
in the range $0.5< z< 2$: $\Delta\alpha/\alpha \siml 10^{-5}$.

\section{$G$}

In this section, we review the experimental constraints on the time variation 
of the gravitational constant. For more detailed earlier review see \citen{gillies97}. 
The results are summarized in Table \ref{table:g}.

\subsection{Planetary motion and $\dot G$}

If we write the effective gravitational constant 
$G$ as $G=G_0+\dot G_0(t-t_0)$,  the effect of changing $G$ is readily seen 
through the change in the equation of motion:
\beq
\frac{d^2{\bf x}}{dt^2}=-\frac{GM{\bf x}}{r^3}=-\frac{G_0M{\bf x}}{r^3}-
\frac{\dot G_0}{G_0}\frac{G_0M}{r}\frac{{\bf x}(t-t_0)}{r^2}.
\eeq
Thus the time variation of $G$ induces an acceleration term of secular type 
in addition to the 
usual Newtonian and relativistic ones, which would affect the motion of 
 bodies, such as planets and binary pulsar. 

A relative distance between the Earth and Mars was accurately measured 
by taking thousands of range measurements between tracking stations 
of the Deep Space Network and Viking launders on Mars.
{}From a least-squares fit of the parameters of the solar system model to 
the data taken from various range measurements including those by 
Viking landers to Mars (from July 1976 to July 1982), a bound on $\dot G$ 
is obtained: $\dot G/G =(2\pm 4)\times 10^{-12}{\rm yr}^{-1}$ 
\citen{hellings83}.

Similarly, Lunar-Laser-Ranging measurements have been used to accurately 
determine parameters of the solar system, in particular the Earth-Moon 
separation. From the analysis of the data from 1969 to 1990, a bound on $\dot G$ is 
obtained: $\dot G/G =(0.1\pm 10.4)\times 10^{-12}{\rm yr}^{-1}$ 
\citen{muller91}; while from the data from 1970 to 1994, 
$\dot G/G =(1\pm 8)\times 10^{-12}{\rm yr}^{-1}$ \citen{williams96}. 
Recent analysis using the data up to April 2004 yields  
$\dot G/G =(4\pm 9)\times 10^{-13}{\rm yr}^{-1}$ \citen{williams04}. 
The uncertainty for $\dot G/G$ is improving rapidly since the sensitivity 
for the observations depends on the square of the time span.

\begin{table}
  \begin{center}
  \setlength{\tabcolsep}{3pt}
  \begin{tabular}{|l|c|c|r|} \hline
  &  redshift & $\Delta G/G$ & $\dot G/G({\rm yr}^{-1})$ \\ \hline
  Viking Lander Ranging\cite{hellings83} & 0 & & $(2\pm 4)\times 10^{-12}$ \\  \hline
  Lunar Laser Ranging\cite{williams04} & 0 & & $(4\pm 9)\times 10^{-13}$ \\  \hline
  Double Neutron Star Binary\cite{taylor91} &  0 & & $(1.10\pm 1.07)\times 10^{-11}$ \\  \hline
  Pulsar-White Dwarf Binary\cite{Verbiest08} &  0 & & $(-5\pm 18)\times 10^{-12}$  \\  \hline
   Helioseismology\cite{krauss98}  & 0  & & $<1.6\times 10^{-12}$ \\  \hline
   White Dwarf Luminosity Function\cite{garcia11}& 0& & $<1.8\times 10^{-12}$ \\  \hline
  Neutron Star Mass\cite{thorsett96} &  $0-3\sim 4$ & & $(-0.6\pm 2.0)\times 10^{-12}$ \\  \hline
  Gravochemical Heating\cite{gravochem} & 0& & $<4\times 10^{-10}$ \\ \hline
  BBN\cite{krauss90} & $10^{9}$ &  $ -0.3\sim 0.4$ & $(-2.9\sim 2.2)\times 10^{-11}$ \\  \hline
  BBN+CMB\cite{krauss04}& $10^{9}$ &  $ -0.15\sim 0.21$ & 
$(-1.5\sim 1.1)\times 10^{-11}$ \\  \hline
  BBN+CMB\cite{cyburt05}& $10^{9}$ &  $ -0.10\sim 0.13$ & 
$(-0.95\sim 0.73)\times 10^{-11}$ \\  \hline
  CMB\cite{ncs2} & $10^3$ &        $< 0.05$          &  $< 3.6 \times 10^{-12}$ \\ \hline
  \end{tabular}
  \end{center}
\caption{
Summary of the experimental bounds on the time variation of the gravitational 
constant. $\Delta G/G\equiv 
(G_{\rm then}-G_{\rm now})/G_{\rm now}$.
}
\label{table:g}
\end{table}

\subsection{Binary Pulsar and $\dot G$}

The timing of the orbital dynamics of binary pulsars provides 
a new test of  time variation of $G$. 
To the Newtonian order, the orbital period of a two-body system is given by
\beq
P_b=2\pi \left(\frac{a^3}{Gm}\right)^{1/2}=
\frac{2\pi \ell^3}{G^2m^2(1-e^2)^{3/2}},
\eeq
where $a$ is the semi-major axis, $\ell=r^2\dot\phi$ is the angular 
momentum per unit mass, $m$ is a Newtonian-order mass parameter, and 
$e$ is the orbital eccentricity. This yields the orbital-period 
evolution rate 
\beq
\frac{\dot P_b}{P_b}=-2\frac{\dot G}{G}+3\frac{\dot\ell}{\ell}-2\frac{\dot m}{m}.
\eeq
Damour, Gibbons and Taylor showed that the appropriate phenomenological 
limit on $\dot G$ is obtained by
\beq
\frac{\dot G}{G}=-\frac{\delta\dot P_b}{2P_b},
\eeq
where $\delta\dot P_b$ represents whatever part of the observed orbital 
period derivative that is not otherwise explained \cite{damour88}. 
{}From the timing of the binary pulsar PSR 1913+16, a bound on $\dot G$ is obtained: 
$\dot G/G= (1.0\pm 2.3)\times 10^{-11}{\rm yr}^{-1}$ \citen{damour88} 
(see also \citen{taylor91}) 
However, only for the orbits of bodies which have negligible gravitational 
self-energies, the simplifications can be made that $\dot P_b/ P_b$ is 
dominated by $-2\dot G/G$ term. When the effect of the variation in 
the gravitational binding energy induced by a change in $G$ is taken into 
account, the above bound is somewhat weakened depending on the equation 
of state \cite{nordtvedt90}. 
This may not be concern to neutron star - white dwarf binaries 
such as PSR B1855+09 ($\dot G/G= (-9\pm 18)\times 10^{-12}{\rm yr}^{-1}$) \citen{taylor94} 
and J0437-4715 ($\dot G/G= (-5\pm 18)\times 10^{-12}{\rm yr}^{-1}$) \citen{Verbiest08}.

\subsection{Stars and $\dot G$}

Since gravity plays an important role in the structure and evolution of 
a star, a star can be a good probe of the time variation of $G$.  
It can be shown that the luminosity 
of a star is roughly proportional to $G^7$ if 
free-free transition dominates the opacity \cite{teller48}. Increasing  
$G$ is effectively the same, via the Poisson equation, as increasing 
the mass or average density of a star, which increases its average mean 
molecular weight and thus increases the luminosity of a star and hence 
decreases its lifetime.  
Since a more luminous star burns more hydrogen, 
the depth of convection zone is affected which is determined directly 
from observations of solar $p$-mode (acoustic wave) spectra \cite{convec91}.
Helioseismology enables us to probe the structure of the solar interior. 
Comparing the $p$-mode oscillation spectra of varying-$G$ solar models with 
the solar $p$-mode frequency observations, a tight bound on $\dot G$ is 
obtained: $|\dot G/G|\leq 1.6\times 10^{-12}{\rm yr}^{-1}$ \citen{krauss98}.

The balance between the Fermi degeneracy pressure of a cold electron gas 
and the gravitational force determines the famous Chandrasekhar mass
\beq
M_{Ch} \simeq G^{-3/2}m_p^{-2},
\eeq
where $m_p$ is the proton mass, which is the upper bound of the masses of  
white dwarfs. White dwarfs are long-lived objects ($\sim 10$ Gyr) and their 
inner cores are almost degenerate,  and hence 
even small variations of $G$ can affect their  structure and evolution. 
Moreover,  white dwarfs do not have nuclear energy sources and their 
energy is of gravitational and thermal origin. The cooling process of 
white dwarfs is now well understood, including the energy release 
due to $~^{22}{\rm Ne}$ sedimentation in the liquid phase and due to 
C/O phase separation on crystallization in the core \cite{garcia10}. 
The decrease in $G$ (larger $G$ in the past) accelerates the cooling 
of white dwarfs.  By comparing the white dwarf luminosity function 
measured in the open cluster NGC 6791 with the simulated luminosity function, 
using the observed distance modulus to break the degeneracy between the age 
of the cluster and the effect of $\dot G$, a tight bound on $\dot G$ is obtained; 
$\dot G/G>-1.8\times 10^{-12}$ \citen{garcia11}.

$M_{Ch}$ sets the mass scale for the late evolutionary stage of 
massive stars, including the formation of neutron stars in core collapse 
of supernovae, and it is thus expected that the average neutron mass is given by 
the Chandrasekhar mass. Measurements of neutron star masses and ages over 
$0<z <3\sim 4$ yield a bound on $\dot G$, 
$\dot G/G=(-0.6\pm 2.0)\times 10^{-12}{\rm yr}^{-1}$ \citen{thorsett96}.

Recently, a new method for constraining $\dot G$ is proposed using 
the surface temperatures of neutron stars (dubbed "gravochemical heating")\cite{gravochem}. 
An increase (or decrease) in $G$ induces the compression (or expansion) of the star. 
Since the chemical potentials depend on the density, the system interior to the star 
departs from the beta equilibrium state, which increases the chemical reaction 
rates so as to reach a new equilibrium state, dissipating energy as internal 
heating and neutrino emission. Comparing the ultrainfrared observation 
of the surface temperature of the millisecond pulsar (PSR J0437-4715), 
upper limits on $\dot G$ are obtained: $|\dot G/G|< 2\times 10^{-10}{\rm yr}^{-1}$  
if direct Urca reaction  operating in the neutron star core is allowed, while  
$|\dot G/G|< 4\times 10^{-10}{\rm yr}^{-1}$ if only modified Urca reactions are 
considered \cite{gravochem}. 

\subsection{Cosmology and $\dot G$: Big-Bang Nucleosynthesis  and 
Cosmic Microwave Background}

\subsubsection{Big Bang Nucleosynthesis.}

The effect of changing $G$ on the primordial light abundances (especially 
$~^4{\rm H_e}$) is already seen in Eq.(\ref{yp}) and Eq.(\ref{freeze}): 
an increase in $G$ increases the expansion rate of the universe, which 
shifts the freeze-out to an earlier epoch and results in a higher abundance 
of $~^4{\rm H_e}$. In terms of the ``speed-up factor''(the ratio of 
the Hubble parameter to that in the Standard Big Bang Nucleosynthesis), 
$\xi\equiv H/H_{SBBN}$,
$Y_p$ is well fitted by \cite{walker91}
\beq
Y_p\simeq 0.244 + 0.074 (\xi^2-1).
\eeq
If $Y_p$ was between $0.22$ and $0.25$, then $-0.32< \Delta G/G < 0.08$, which 
corresponds to $\dot G/G=(-0.55\sim 2.2)\times 10^{-11}{\rm yr}^{-1}$. 
A similar (more conservative) bound was obtained 
in \citen{krauss90} : $-0.3< \Delta G/G < 0.4$.
Combining the determination by WMAP of $\Omega_Bh^2$
 and recent measurements of primordial 
deuterium abundance (but without Helium and Lithium abundance), 
a slightly tighter constraint is obtained \cite{krauss04} : 
$-0.15<\Delta G/G<0.21$. By combining WMAP value of $\Omega_Bh^2$ and 
recent results of the reanalysis of helium abundance \cite{olive04}, 
a similar bound has been obtained: 
$-0.10<\Delta G/G<0.13$ \citen{cyburt05}. It should be noted however that 
the analysis by WMAP team assumed the Einstein gravity and the effect 
of changing $G$ is not included to determine $\Omega_Bh^2$. 
Hence, these analyses are not consistent and 
should be made in the context of scalar-tensor gravity (or its variants) consistently. 

\subsubsection{Cosmic Microwave Background.}

Changing $G$ changes the Hubble parameter and hence changes 
the size of horizon: $H^{-1}\propto G^{-1/2}$, which results in the change 
of both the location and the amplitude of acoustic peaks through the projection effects and 
the diffusion damping scale \cite{ncs}. For example, an increase in $G$ shifts 
the peak positions in the spectrum toward higher values of $\ell$. 
A larger $G$ makes the diffusion length $\lambda_D \simeq 1/\sqrt{H\dot\tau}$ 
shorter and thus weakens the diffusion damping at the peak positions 
because the peak positions also depend on $H^{-1}$ ($\lambda\propto H^{-1}$) and hence 
the damping factor $\exp(-\lambda_D^2/\lambda^2)$ decreases less.

Recently, anisotropies in the cosmic microwave background have been 
measured up to $\ell < 800$ by WMAP satellite \cite{wmap}. 
{}From the analysis of the WMAP data within the context varying $G$ model, 
the variation of the gravitational constant at the recombination epoch is 
constrained as \cite{ncs2},  $\Delta G/G < 0.05$. 

\begin{figure}
\includegraphics[width=13cm]{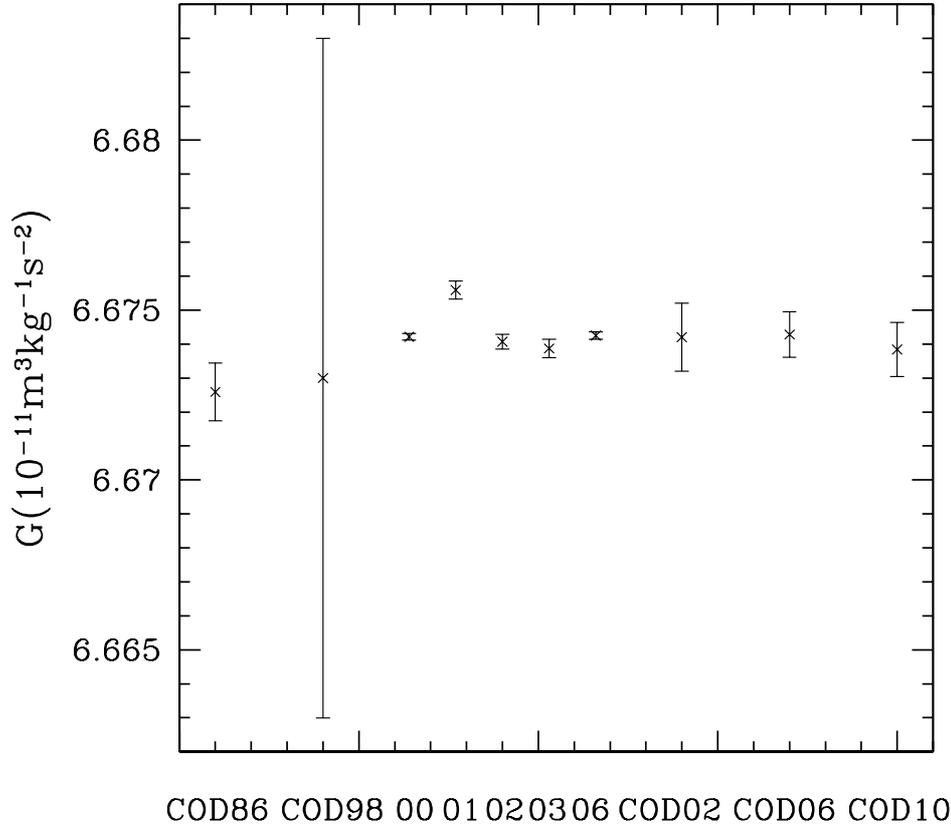}
\caption{Experimental results on the measurements of $G$ and the CODATA recommended value of $G$.}
\label{fig:g}
\end{figure}

\subsection{Measuring $G_0$: Recent Developments}

In 1798, Cavendish carried out experiments to measure $G_0$ by 
using a torsion balance apparatus (proposed and constructed by John Michell),  
which has become known as the Cavendish Experiment \cite{cavendish}. 
The method is still used basically in measuring $G_0$. 

The torsion balance consists of a dumbbell suspended from the middle 
by a thin fiber. The dumbbell consists of two small masses ($m$) fastened 
to a thin rod of length $2\ell$. When a large pair of masses ($M$) are brought 
into proximity to the smaller masses, the dumbbell rotates by an angle $\varphi_0$ 
and comes to a halt.  
{}From the balance between the torsion of the fiber (the torsion coefficient $D$) 
and the torque due to the gravity force between $m$ and $M$ (the distance $r$), 
we have $D\varphi_0=2G_0Mm\ell/r^2$. When the large masses are removed from the set up, 
the dumbbell begins to oscillate because of the restoring force of the fiber. 
The period of the oscillation, $T$, is given by $T=2\pi\sqrt{I/D}$, where $I(=2m\ell^2)$ 
is the moment of inertia of the dumbbell. Eliminating $D$, $G_0$ is thus determined by 
$G=4\pi^2 r^2 \varphi_0\ell/MT^2$. 

No laboratory measurements of $\dot G/G$ has been performed recently (see 
\citen{gillies97} for older laboratory experiments). This is mainly 
because the measurements of the present gravitational constant $G_0$ itself 
suffer from systematic uncertainties and have not been performed with good 
precision. For example, due to the uncertainties associated with the frequency 
dependence of the torsion coefficient of fibers, the recommended value of $G_0$ 
by CODATA (Committee on Data for Science and Technology) became worsened 
from $G=(6.67259 \pm 8.5\times 10^{-4})\times 10^{-11} {\rm m}^3{\rm kg}^{-1}
{\rm s}^{-2}$ in 1986 to $(6.673\pm 1.0\times 10^{-2})\times 10^{-11} $ in 1998 \cite{codata1998}.

Gundlach and Merkowitz measured $G_0$ with a torsion-balance 
experiment in which string-twisting bias was carefully eliminated 
\cite{gundlach00}. The result was a value of 
$G_0=(6.674215\pm 0.000092)\times 10^{-11}{\rm m^3kg^{-1}s^{-2}}$.
Recently, however, the measurement of $G$ 
with a torsion-strip balance resulted in 
$G_0=(6.67559\pm 0.00027)\times 10^{-11}{\rm m^3kg^{-1}s^{-2}}$,
 which is 2 parts in $10^4$ higher than the 
result of Gundlach and Merkowitz \cite{quinn01}. Probably the difference is 
still due to systematic errors hidden in one or both of the measurements. 
On the other hand, recent beam balance measurement 
($G_0=(6.67407\pm 0.00022)\times 10^{-11}{\rm m^3kg^{-1}s^{-2}}$) \citen{sch02}
and torsion balance measurement ($G_0=(6.67387 \pm 0.00027)\times 
10^{-11}{\rm m^3kg^{-1}s^{-2}}$) \citen{arm03}
of $G$ are consistent with the result by Gundlach and Merkowitz. 
A new measurement of $G$ was made using a beam balance \cite{schl06}. 
The measured value, $G_0=(6.674252 \pm 0.000109)\times 10^{-11}{\rm m^3kg^{-1}s^{-2}}$, is 
consistent with that by Gundlach and Merkowitz.
\footnote{However, 
two recent determinations of $G$, one by comparing the time it took for a torsion pendulum 
to swing past masses placed at varying distances from it \cite{luo09} 
and another by using a laser interferometer to measure 
the displacement of pendulum bobs by various masses \cite{parks10}, give values significantly 
deviated from the value by Gundlach and Merkowitz: 
$G_0=(6.67349 \pm 0.00018)\times 10^{-11}{\rm m^3kg^{-1}s^{-2}}$ \citen{luo09} and 
$G_0=(6.67259 \pm 0.00085)\times 10^{-11}{\rm m^3kg^{-1}s^{-2}}$ \citen{parks10}. 
The source of the inconsistency is currently unknown, and these new values may make   
the next recommended value of $G$ by CODATA decrease and make the uncertainty larger.  } 
The recommended value of $G$ by 
CODATA is being improved: $G=(6.6742\pm   1.0\times 10^{-3})\times 10^{-11} 
{\rm m}^3{\rm kg}^{-1}{\rm s}^{-2}$ for CODATA 2002 \cite{codata2002}, while  
$G=(6.67428\pm   6.7\times 10^{-4})\times 10^{-11} 
{\rm m}^3{\rm kg}^{-1}{\rm s}^{-2}$ for CODATA 2006 \cite{codata2006}. \footnote{The 2010 CODATA recommended value of $G$ is 
$G=(6.67384\pm   8.0\times 10^{-4})\times 10^{-11} 
{\rm m}^3{\rm kg}^{-1}{\rm s}^{-2}$.\cite{nist}}
The values of $G$ as a function of the year of measurements are shown in Fig. \ref{fig:g}. 

Moreover, conceptually different experiments of measuring $G$ using a gravity gradiometer 
based on cold-atom interferometry were performed \cite{fixler07,lamporesi08}. 
Freely falling samples of 
laser-cooled  atoms are used in a gravity gradiometer to probe the field generated by nearby 
source masses. A measured value of $G$ is, $G_0=(6.667\pm   0.011\pm0.003)\times 10^{-11} 
{\rm m}^3{\rm kg}^{-1}{\rm s}^{-2}$ \citen{lamporesi08}. It may be possible to push the measurement 
accuracy below $10^{-4}$.

As the accuracy of the measurements improves, it may be possible to 
 place a bound on the present-day variation of $G$. 
Although the current accuracy of G is more than 6 digits worse than the solar system experiments or 
cosmological constraints to place a constraint on the present-day $\dot G$, the situation will be changed 
after a century since the accuracy improves by one digit during these ten years. 
It is important to 
pursue laboratory measurements of $\dot G/G$ since they 
are repeatable and hence are complementary to astrophysical and geophysical 
constraints.

\section{Proton-Electron Mass Ratio}

In this section, we briefly mention the experimental constraints on the variation 
of the 
electron-proton mass ratio, $\mu=m_p/m_e$. 
The results are summarized in Table \ref{table:pe}.

\begin{table}
  \begin{center}
  \setlength{\tabcolsep}{3pt}
  \begin{tabular}{|l|c|c|r|} \hline
  &    redshift & $\Delta\mu/\mu$ &  $\dot\mu/\mu({\rm yr}^{-1})$ \\ \hline
  Atomic Clock(Mg)\cite{godone93} & 0 &           & $(2.5\pm 2.3)\times 10^{-13}$\\ \hline
  Atomic Clock(Hg)\cite{bize03} & 0 &             & $\pm 7.0\times 10^{-15}$ \\  \hline
Atomic Clock(Sr)\cite{blatt08} & 0&                &$(-1.6\pm 1.7)\times 10^{-15}$ \\  \hline
Molecular Clock(${\rm SF_6}$)\cite{shelkovnikov08} & 0&  &  $ (-3.8\pm 5.6)\times 10^{-14}$ \\ \hline
  HI \cite{cowie95} & 2.811 & $(-0.8\pm 3.1)\times 10^{-4}$ & 
$(0.7\pm 2.7)\times 10^{-14}$ \\  \hline
${\rm H_2}$ \cite{cowie95} & 1.7764 & $(-0.7\pm 0.6)\times 10^{-4}$ & 
$(0.7\pm 0.6)\times 10^{-14}$ \\  \hline
 HI/${\rm H_2}$\cite{tza05} & $0.24-2.04$ & $(-1.29\pm 1.01)\times 10^{-5}$ &  \\  \hline
  ${\rm H_2}$ \cite{potekhin98} & 2.81 & $(8.3^{+6.6}_{-5.0})\times10^{-5}$ & 
$(-7.3^{+4.5}_{-5.8})\times 10^{-15}$ \\  \hline
   ${\rm H_2}$ \cite{ivanchik02} & 2.3377, 3.0249 & $(5.7\pm 3.8)\times 10^{-5}$ & 
$(-5.1\pm 3.4)\times 10^{-15}$ \\  \hline
   ${\rm H_2}$ \cite{ubachs04} & 2.3377, 2.8108, 3.0249 & $(-0.5\pm 3.6)\times 10^{-5}$ & 
$0.4\pm 3.2)\times 10^{-15}$ \\  \hline
   ${\rm H_2}$ \cite{ivanchik05} & 2.5947, 3.0249 & $(1.65\pm 0.74)\times 10^{-5}$ & 
$(-1.46\pm 0.65)\times 10^{-15}$ \\  \hline
  ${\rm H_2}$ \cite{reinhold06} & 2.5947, 3.0249 & $(2.44\pm 0.59)\times 10^{-5}$ & 
$(-2.16\pm 0.52)\times 10^{-15}$ \\  \hline
${\rm H_2}$ \cite{king08} & 2.595, 3.025, 2.811 & $(2.6\pm 3.0)\times 10^{-6}$ & 
$(-2.3\pm 2.7)\times 10^{-16}$ \\  \hline
  ${\rm H_2/HD}$ \cite{malec10} & 2.059 & $(5.6\pm 5.5)\times 10^{-6}$ & 
$(-5.3\pm 5.2)\times 10^{-16}$ \\  \hline
  ${\rm H_2/HD}$ \cite{weerdenburg11} & 2.059 & $(8.5\pm 3.6)\times 10^{-6}$ & 
$(-8.0\pm 3.4)\times 10^{-16}$ \\  \hline
 ${\rm H_2/HD}$ \cite{king11} & 2.811 & $(0.3\pm 3.2)\times 10^{-6}$ & 
$(-0.3\pm 2.8)\times 10^{-16}$ \\  \hline
   OH/${\rm HCO^+}$/HI \cite{kanekar03} & 0.684 & $(0.27\pm 1.6)\times 10^{-3}$ & 
$(-0.44\pm 2.6)\times 10^{-13}$ \\ \hline 
   OH \cite{kanekar10} & 0.247 &  $(-6.2\pm 2.4)\times 10^{-6}$ &  
$(2.1\pm 0.8)\times 10^{-15}$\\ \hline
   OH/HI \cite{kanekar05} & 0.765, 0.685 &  $\pm 1.4\times 10^{-5}$ & 
$<2.2\times 10^{-15}$\\ \hline
${\rm NH_3}$/CO,${\rm HCO^+}$,HCN \cite{flambaum07} & 0.68466 & $(0.6\pm 1.9)\times 10^{-6}$ & 
$(-0.9\pm 3.0)\times 10^{-16}$\\ \hline
${\rm NH_3}$/${\rm HCO^+}$,HCN \cite{murphy08} & 0.68466 & $(0.74\pm 0.47)\times 10^{-6}$ &
$(-1.2\pm 0.74)\times 10^{-16}$\\ \hline
${\rm NH_3}$/CS, ${\rm H_2CO}$ \cite{kanekar11} & 0.685 & $(-3.5\pm 1.2)\times 10^{-7}$ &
$(5.5\pm 1.9)\times 10^{-17}$\\ \hline
${\rm NH_3},{\rm HC_3N}$ \cite{henkel09} & 0.89 & $(0.08\pm 0.47)\times 10^{-6}$ & 
$(-0.11\pm 0.64)\times 10^{-16}$\\ \hline
    \end{tabular}
  \end{center}
\caption{
Summary of the experimental bounds on the time variation of the mass ratio of electron and proton, 
$\mu=m_p/m_e$. $\Delta\mu/\mu\equiv 
(\mu_{\rm then}-\mu_{\rm now})/\mu_{\rm now}$.}
\label{table:pe}
\end{table}

\subsection{Molecular lines and $\mu$}

Thompson noted the different dependence of the electronic, vibrational, and rotational 
energy levels on $\mu$ and first pointed out the possibility that 
the presence of cosmological evolution in $\mu$ can be tested by using 
observations of molecular hydrogen in quasar absorption systems \cite{thompson75}. 
In the Born-Oppenheimer approximation, the molecular hydrogen levels depend 
on $m_p$ and can be written as
\beq
E=E_{elec}+\frac{E_{vib}}{\sqrt{\mu}}+\frac{E_{rot}}{\mu}.
\eeq 
Thus, the energy shift in any vibration-rotation transition $j$ 
in the Lyman series has the form
\beq
\Delta E_j=a_{elec}+b_j/\sqrt{\mu}+c_j/\mu,
\eeq
and the difference in energy between two transitions is 
\beq
\Delta E_i-\Delta E_j\simeq b_{ij}/\sqrt{\mu}+c_{ij}/\mu.
\eeq
Hence, to lowest order, a change in $\mu$ induces a change in $\Delta E_i-\Delta E_j$: 
\beq
\frac{\delta \mu}{\mu}\simeq -2\frac{\delta  (\Delta E_i-\Delta E_j)}{ \Delta E_i-\Delta E_j}
\simeq -\frac{\delta v}{ c}\frac{2\Delta E_i}{ \Delta E_i-\Delta E_j},
\eeq
where $\delta v $ is the mean offset, compared to the laboratory value, of the energy 
difference between the two sets of lines, when that 
offset is represented as a velocity difference. 
Based on this method, Pagel first obtained a constraint on $\dot\mu$, 
$|\dot\mu/\mu|<5\times 10^{-11}{\rm yr}^{-1}$ from the comparison of 
different redshifts determined by neutral hydrogen molecule and by heavy ion absorption lines 
\cite{pagel77}. 
The ratio of the hyperfine 21 cm 
absorption transition of neutral hydrogen to an optical resonance transition is dependent on 
$g_p\mu\alpha^2$, where $g_p$ is the proton gyromagnetic ratio. 
Comparing the measured redshifts of 21 cm 
and optical absorption a constraint on the change in $\mu$ is obtained at $z=1.7764$ \citen{cowie95}: 
$\Delta\mu/\mu=(-0.7 \pm 0.6)\times 10^{-4}$ 
assuming that $\alpha$ and $g_p$ are constants (note the different definition of $\mu$ used there 
and we changed the 95\% confidence limits given in \citen{cowie95} into 1$\sigma$ ones). 
Various observational constraints obtained so far are summarized in Table \ref{table:pe}. 

Recent measurements of ${\rm H_2}$ lines of Lyman and Werner bands at 
$z=2.5947$ and 3.0249 toward the quasars 
Q0405-443 and Q0347-383 with VLT/UVES indicate a systematic 
shift of $\mu$ in the past, $\Delta\mu/\mu=(1.65\pm0.74)\times 10^{-5}$ \citen{ivanchik05}, 
but the results depend on the laboratory wavelengths:  the above value is for wavelengths 
derived from a direct determination using laser techniques, 
while $\Delta\mu/\mu=(3.05\pm0.75)\times 10^{-5}$ 
for those derived from energy level determination \cite{ivanchik05}.  In general, 
a measured $i$-th molecular line wavelength $\lambda_i$ in an absorption system at 
redshift $z_{abs}$ is given by \cite{potekhin98}
\beqa
\lambda_i=\lambda_i^0(1+z_{abs})(1+K_i\Delta\mu/\mu),
\eeqa
where $\lambda_i^0$ is the laboratory transition wavelength and 
$K_i=d\ln \lambda_i/d\ln \mu$ is the sensitivity coefficient. 
With an improved calculation of sensitivity coefficients $K_i$ and 
new accurate laboratory spectroscopic measurements, the 
reanalysis of the data \cite{ivanchik05} strengthens the case for 
a larger $\mu$ in the past \cite{reinhold06}: 
$\Delta\mu/\mu=(2.44\pm0.59)\times 10^{-5}$. 
However, it is pointed out that the techniques used to calibrate the wavelength scale of the 
Ultraviolet and Visual Echelle Spectrograph (UVES) on VLT produce calibration errors \cite{murphy07}. 
Reanalysis of the spectra by using the improved wavelength calibration techniques and improved 
fitting procedures yields a constraint on 
$\Delta\mu/\mu$ as $\Delta\mu/\mu=(2.6\pm 3.0)\times 10^{-6}$, 
which is consistent with no variation at 1 $\sigma$ \citen{king08}.  
Recent observations of a number of hydrogen lines (${\rm H_2}$ and HD) 
in the spectrum of J2123-0050 with Keck/HIRES at $z=2.059$ yield a constraint on $\Delta\mu$: 
$\Delta\mu/\mu=(5.6\pm 5.5({\rm stat})\pm 2.9({\rm sys}))\times 10^{-6}$ \citen{malec10}.
The analysis of the spectrum of the same object observed with VLT/UVES gives a similar constraint: 
$\Delta\mu/\mu=(8.5\pm 3.6({\rm stat})\pm 2.2({\rm sys}))\times 10^{-6}$ \citen{weerdenburg11}. 
The analysis of a new spectrum of Q0528-250 with VLT/UVES yields 
$\Delta\mu/\mu=(0.3\pm 3.2({\rm stat})\pm 1.9({\rm sys}))\times 10^{-6}$ \citen{king11}.

Recently, the method of using 
18 cm OH lines has been proposed \cite{kanekar03,darling03,darling04} 
to avoid possible systematic errors from multiple 
species which may have systematic velocity offsets. The ground  
$~^2\Pi_{3/2}J=3/2$ rotation state of OH is split into two levels by 
$\Lambda$ doubling and each of these $\Lambda$ doubling states is 
further split into two hyperfine states. Transitions between these levels lead to 
four spectral lines with wavelength $\sim$ 18 cm. Transitions with $\Delta F=0$ 
are called the main lines, with frequencies of 1665.4018 and 1667.3590 MHz, while 
transitions with $\Delta F=1$ are called satellite lines, with frequencies of 
1612.2310 and 1720.5299 MHz. Since the four OH lines arise from two very different 
physical processes, $\Lambda$-doubling and hyperfine splitting, the transition 
frequencies have different dependences on the fundamental constants, 
$\alpha$ and $\mu$ and the  proton gyromagnetic ratio $g_p$. 
Therefore, measurements of these lines enable  
us to constrain variations in $\alpha$ and $\mu$ from a single species. 
The radio observations of two satellite lines at $z=0.247$ yield 
$\Delta X/X=(2.2\pm 3.8)\times 10^{-5}$ for $X=g_p(\mu\alpha^2)^{1.85}$ \citen{kanekar04}. 
Assuming that the variations of $\alpha$ and the proton gyromagnetic ratio are small, 
a change in $\mu$ is constrained as, 
$\Delta\mu/\mu=(1.2\pm 2.0)\times 10^{-5}$ \citen{kanekar04}. 
Deep Westerbork Synthesis Radio Telescope and Arecibo Telescope observations of these 
satellite lines yield  
$\Delta X/X=(-1.18\pm 0.46)\times 10^{-5}$, suggesting $2.6\sigma$ evidence for a change in $X$ 
\citen{kanekar10}. The limiting cases, assuming that only $\alpha$ or $\mu$ changes, 
are $\Delta\alpha/\alpha=(-3.1\pm 1.2)\times 10^{-6}$ and 
$\Delta\mu/\mu=(-6.2\pm 2.4)\times 10^{-6}$ \citen{kanekar10}.   
All four 18 cm OH lines have  
recently been detected at $z=0.765$ with low signal-to-noise ratio \cite{kanekar05}, 
which, when combined with HI 21 cm lines at $z=0.685$, yields a constraint of 
$|\Delta\mu/\mu|<1.4\times 10^{-5}$ \citen{kanekar05}.

It is pointed out recently that the inversion transition frequencies of ammonia (${\rm NH_3}$) are 
significantly sensitive to the variation of $\mu$ \cite{flambaum07}. By comparing the inversion 
spectrum of ${\rm NH_3}$ at $z=0.6847$ (toward B0218+357) 
with rotational spectra of other molecules (CO,${\rm HCO^+}$,HCN), a strong constraint is obtained: $\Delta\mu/\mu=(-0.6\pm 1.9)\times 10^{-6}$ 
\citen{flambaum07}. More detailed comparison of the ${\rm NH_3}$ inversion transitions 
with ${\rm HCO^+}$ and HCN molecular rotational transitions gives a stronger constraint on $\Delta\mu$:  
$\Delta\mu/\mu=(0.74\pm 0.47({\rm stat})\pm 0.76({\rm sys}))\times 10^{-6}$ and a 2$\sigma$ 
constraint of  $|\Delta\mu/\mu|<1.8\times 10^{-6}$ \citen{murphy08}. 
The bound is recently updated further by comparing the ${\rm NH_3}$ inversion lines  with molecular rotational lines (CS, ${\rm H_2CO}$)  
by using the Green Bank Telescope:  $\Delta\mu/\mu=(-3.5\pm 1.2)\times 10^{-7}$ \cite{kanekar11}. 
Also, from the comparison of the ${\rm NH_3}$ inversion transitions 
with ${\rm HC_3N}$ molecular rotational transitions at $z=0.89$ observed with the Effelsberg radio telescope, a similar bound 
is obtained: $\Delta\mu/\mu=(0.08\pm 0.47)\times 10^{-6}$ and a 3$\sigma$ 
constraint of  $|\Delta\mu/\mu|<1.4\times 10^{-6}$ \citen{henkel09}. 
Moreover, from the spectral observations of molecular cores in the disk of 
the Milky Way in molecular transitions of ${\rm NH_3}$ and ${\rm HC_3N}$  at the Effelsberg radio telescope, 
a statistically significant velocity offset 
$23\pm 4({\rm stat})\pm 3({\rm sys}){\rm m/s}$ between the radial velocities 
 ${\rm NH_3}$ and ${\rm HC_3N}$ is found \cite{levshakov10}. 
When interpreted in terms of the local (spatial) variation of $\mu$, 
this implies a tentative signal of 
$\Delta\mu/\mu=(-26\pm 1({\rm stat})\pm 3({\rm sys}))\times 10^{-9}$ \citen{levshakov10}, where 
$\Delta\mu\equiv (\mu_{{\rm MilkyWay}}-\mu_{{\rm lab}})/\mu_{{\rm lab}}$. 
However, since the number of sources is small and a different velocity offset 
is observed at a different epoch of the observations, there may exist 
unaccounted-for systematic effects. 

Very recently, torsion-vibrational frequencies of methanol (${\rm CH_3OH}$) are found to be 
far more sensitive to the  variation of $\mu$ \cite{jansen11}. Using the published data of observing narrow emission lines of the methanol masers in the Milky Way, 
the local (spatial) variation of $\mu$ is constrained as 
$\Delta\mu/\mu=(-11\pm17)\times 10^{-9}$ \citen{levshakov11}. 

\subsection{Laboratory Tests: Clock Comparison}

Laboratory limits on the variations of $\mu$ are also obtained by comparison of atomic clocks \cite{godone93,bize03,blatt08} as explained in Sec. 2.3. 
As for the constraint using molecular clocks, 
recently, from the comparison of the frequency of a rovibrational 
transition in ${\rm SF_6}$ with the hyperfine transition in Cs, combined with \citen{fortier07} 
to break the degeneracy with variations of $\alpha$ and nuclear magnetic moment, 
a constraint of $\dot\mu/\mu=(-3.8\pm 5.6)\times 10^{-14}{\rm yr}^{-1}$  
is obtained \cite{shelkovnikov08}.

\section{$\Lambda$ or Dark Energy}

Finally, we briefly comment on the potential variability of the 
cosmological constant (or dark energy) because in the runaway scenario of 
dilaton or moduli $\phi$, $\dot\alpha/\alpha$ and 
$\dot G/G$ would close to $\dot\phi/\phi$ \citen{witten00}. 

\subsection{Evidence for $\Lambda >0$}

There are two arguments for the presence of dark energy. The first indirect 
evidence comes from the sum rule in cosmology:
\beq
\sum \Omega_i=1,
\eeq
where $\Omega_i\equiv 8\pi G\rho_i/3H^2_0$ is the density parameter of the 
i-th energy component, $\rho_i$. The density parameter of the curvature, 
$\Omega_K$, is defined by  $\Omega_K\equiv -k/a^2H^2_0$. Since the current 
observational data indicate that matter density is much less than the 
critical density $\Omega_M<1$ and that the Universe is flat, we are led to 
conclude that the Universe is dominated by dark energy, 
$\Omega_{DE}=1-\Omega_M-\Omega_K>0$.

The second evidence for dark energy is from the observational evidence for the
accelerating universe \cite{sn} \footnote{The Nobel prize in physics 2011 is awarded to 
S. Perlmutter, B. P. Schmidt and A. G. Riess
 for the discovery of the accelerating expansion of the Universe through observations of distant supernovae. Congratulations !}:
\beq
\frac{\ddot a}{ aH_0^2}=-\frac12\left(\Omega_M(1+z)^3+
(1+3w)\Omega_{DE}(1+z)^{3(1+w)}\right)>0,
\eeq
where $w$ is the equation of state of dark energy, $w\equiv p_{DE}/\rho_{DE}$.
Since distance measurements to SN\,Ia strongly indicate the Universe is 
currently accelerating, the Universe should be dominated by dark energy with 
negative pressure ($w<0$). We note that another argument for negative 
pressure comes from the necessity of the epoch of the matter domination.

\subsection{Supernova and $\dot\Lambda$}

A current bound on the equation of state of dark energy 
from supernova data (580 supernovae) is $|w-1|\siml 0.07$ \citen{suzuki11}.
Future observations of high redshift supernovae/galaxies/clusters/BAO  
would pin down the bound on $w$ to $|w-1|\siml 0.01$.
The extent of  the time variation of dark energy density is readily seen from 
the equation of motion:
\beq
\frac{\dot\rho_{DE}}{\rho_{DE}}=-3(1+w)H.
\eeq

\section{Conclusion}

A short account of the experimental constraints on the time variability of 
the constants of nature ($\alpha$, $G$ and $\mu$) was given. 
Since there are some theoretical motivations for the time variability of 
the constants of nature and the implications of it are profound, 
it is worth examining whether the constancy of the constants of nature 
is just a very good approximation. 

Let's keep shaking the pillars to make sure they're rigid! \cite{nature}

~

\section*{Acknowledgements}
The author would like to acknowledge useful discussions with Yasunori Fujii, Naoto Kobayashi, 
Masahide Yamaguchi and Jun'ichi Yokoyama at various stages in completing this review. 
This work was supported in part by a Grant-in-Aid for Scientific 
Research (Nos.20540280 and 15740152) from the Japan Society for the Promotion of
Science.


\end{document}